# Magnetic properties of iron nanoparticles prepared by exploding wire technique

Abdullah Alqudami and S Annapoorni [*]

Department of Physics and Astrophysics, University of Delhi, Delhi, India 110 007

Subhalakshmi Lamba

School of Sciences, Indira Gandhi National Open University,

Maidan Garhi, New Delhi, India 110068

P C Kothari and R K Kotnala

National Physical Laboratory, Dr. K. S. Krishnan Marg, New Delhi, India 110 012

**Abstract**

Nanoparticles of iron were prepared in distilled water using very thin iron wires and sheets, by the electro-exploding wire technique. Transmission electron microscopy reveals the size of the nanoparticles to be in the range 10 to 50 nm. However, particles of different sizes can be segregated by using ultrahigh centrifuge. X-ray diffraction studies confirm the presence of the cubic phase of iron. These iron nanoparticles were found to exhibit fluorescence in the visible region in contrast to the normal bulk material. The room temperature hysteresis measurements upto a field of 1.0 tesla were performed on a suspension of iron particles in the solution as well as in the powders obtained by filtration. The hysteresis loops indicate that the particles are superparamagnetic in nature. The saturation magnetizations was ~ 60 emu / gm. As these iron particles are very sensitive to oxygen a coating of non-magnetic iron oxide tends to form around the particles giving it a core - shell structure. The core particle size is estimated theoretically from the magnetization measurements. Suspensions of iron nanoparticles in water have been proposed to be used as an effective decontaminant for ground water.

[*] Corresponding author: Email: annapoorni@physics.du.ac.in
Tel. +9111-27662276

## 1. Introduction

Study of magnetic properties of iron nanoparticles can be applied to different branches of science and technology, allowing interesting information to be obtained. With decreasing particle size, the fraction of atoms lying on the surface and interfacial regions increases, making the effect of surface/interface electronic structure on the magnetic properties more important. The magnetic properties of the surface atoms are decided by the number of magnetic neighbor atoms and are also affected by chemisorption. Bodker et al. showed that the chemisorption of oxygen on $\alpha$-Fe particles results in a large surface hyperfine field [1, 2]. Furthermore, it is known that both the coercivity and the effective magnetic anisotropy decrease with the thickening of surface oxide layer [3]. The physics of magnetic clusters exposed on surfaces or embedded into matrices has been reviewed in detail by Bansmann et al. [4]. The high surface activity gives iron nanoparticles the opportunity to represent a new generation of environmental remediation technologies [5].

Equally important, the preparation methods influence the properties of nanoscale iron particles. Mostly chemical and physical methods are used to generate iron nanoparticles in the form of colloids as well as powders [6-10]. From the preparation point of view, one of the important subjects in the study of magnetic nanoparticles is the preparation of stable nanoparticles with novel properties. In the present study we report the characterization and unusual optical and magnetic properties of iron nanoparticles prepared using the exploding wire technique [11, 12]. Poly-vinyl pyrrolidone (PVP) polymers have been used to stabilize the nanoparticles in a water medium. Both water- based and powders of the iron nanoparticles were used to investigate their properties.  The core radius of the iron nanoparticles was calculated from a theoretical model, on the basis of their magnetic properties.  To explain the concentration dependent properties we present the results of numerical simulation studies of the hystersis using a random anisotropy, interacting model  of single domain magnetic particles.

## 2. Experimental

Iron wires  (99.998 %) having a diameter of 0.2 mm were exploded in double distilled water and double distilled water containing 25, 50 and 100 µM poly-vinyl pyrrolidone (PVP) polymer ($M_n$ = 40,000, Aldrich) to produce iron nanoparticles and PVP coated iron nanoparticles respectively. These wires have been exploded by bringing the wire into sudden contact with iron plate (purity

99.998 %) when subjected to a potential difference of 36 V DC. The process involves the generation of high current density throughout thin wire, which causes the fragmentation of wire to very small clusters. The process involves the generation of plasma as well. Van der Waals attraction forces then give rise to the formation of nanosized particles. The strong affinity of iron towards oxygen may lead to the formation of oxide layers around the nanoparticles. The presence of PVP polymers in different concentrations provides another kind of passivation.

The total mass being exploded was about 0.3 gm in all the solutions. The explosions in all the experiments were carried out with the same conditions.

Iron nanoparticles in the dried powder form were used for the x-ray diffraction studies that were performed with a Philips Analytical X-Ray Diffractometer type PW3710 Based using Cu-K$\alpha$ radiation (wavelength 1.54056 Å). Small drops from the decanted solutions were allowed to dry on carbon-coated copper grid for the electron microscopy imaging using JEOL JEM 2000EX transmission electron microscopy (TEM). The UV-Visible absorption spectra for the water-based solutions containing colloidal iron nanoparticles have been recorded by using UV-2510PC spectrophotometer and the fluorescence spectra were recorded using Edinburgh Analytical Time Resolved Fluorometer. The magnetic moment (M) of iron nanoparticles in the powder form and the water-based iron nanoparticles was measured using Lake Shore 7304 Vibrating Sample Magnetometer (VSM). Thermal analyses were performed on powder iron nanoparticles by using Shimadzu DTG-60 thermogravimetric – differential thermal analysis (TG-DTA).

### 3. Results and discussion

#### I. X-ray diffraction and transmission electron microscopy

Figure 1 shows the X-ray diffraction of the iron nanoparticles in reference to bulk iron wire. The peak obtained around 44.5 degree corresponds to the (110) line of (FCC) $\alpha$ ? Fe. No oxide peak was observed. The lattice constant obtained is in agreement with the literature value. TEM images of both iron nanoparticles and PVP-coated iron nanoparticles are shown in figure 2. Fig. 2(a) shows large spherical clusters, where the nanoparticles were obtained in pure water. From the diffraction patterns (inset in Fig. 1(a)), d-spacing values correspond to the $\alpha$– Fe phase and iron oxide phase were obtained. We believe that the exploded wires will result in small clusters (few atoms) with high filling factor and this will ultimately lead to the formation of cluster aggregates. This aggregation will saturate when the filling becomes less. However, some amounts of stabilizers such as PVP polymer can terminate the aggregation as seen in figure 2(b), where small nanoparticles with mean diameter of about 10 nm are observed. Water as a medium will provide fast cooling for the generated particles with energetically favored shapes. Moreover, the high surface area per unit mass of the resulted iron nanoparticles will enhance the surface activity and tend to react with water molecules. Iron nanoparticle surface, in the absence of polymers, thus may get oxidized. This process results in forming some sort of oxide layer. The coating process occurs during and after the particle growth and agglomerations. The oxidation of iron nanoparticles is reflected through the increasing of pH from 6.4 for the distilled water used to 7.14 for the water-based iron nanoparticles solution. PVP coated or stabilized iron nanoparticles solution has a pH value of 6.82.

### II. UV-visible and Fluorescence

Iron nanoparticles obtained using the exploding wire showed ultraviolet absorptions and sufficient amount of visible fluorescence (figure 3). The absorption spectra agreed with the extinction generated using Mie's theory. Details of the optical properties of iron nanoparticles are communicated elsewhere. The fluorescence behavior of iron nanoparticles is explained as transitions among the surface/interface electronic energy states, which is totally a surface phenomenon.

### III. Magnetization measurements

The present study focuses on the magnetic properties of these novel iron nanoparticles. Figure 4 shows the magnetization measurements of iron nanoparticles coated with different amount of PVP. The measurements were performed on powder samples. It is observed that the saturation magnetization does not exceed 60 emu / gm for pure iron nanoparticles while the magnetization values for coated iron nanoparticles are much less. In the graphs we mention the mass relatives between the mass of iron exploded and the mass of PVP added to the explosion mediums as 3:1, 3:2 and 3:4 (Fe: PVP), which correspond to 25, 50 and 100 µM of PVP respectively. The low value of the coercivity also indicates that the system is close to the superparamagnetic transition at room temperature. The suppression of the magnetization of PVP-coated samples relative to the pure iron nanoparticles is obviously partly due to mass effects as a result of the coating of the polymer on the nanoparticles and can give us a quantitative estimate of the degree of polymer coating on the different samples. For this purpose the thermal analyses for one of the PVP-coated samples (which is Fe: PVP = 3:1) and for the pure PVP polymer have been carried out and the results are shown in figure 5. The weight loss was about 23% for the PVP-coated iron nanoparticles while it was more than 95% for the pure PVP, which is in agreement with that reported in literatures [13, 14]. Using this result it is possible to make a rough estimate of the PVP content in the remaining compositions, with which one can make the corrections for the mass effect in the composite samples.

However when the correction for the mass effects is made to the magnetization data, the corresponding values for the saturation magnetization in the different compositions does increase substantially but it still falls short of the values measured for the pure iron nanoparticles (figure 7). In addition this effect increases with increasing polymer content.

It is our belief that the presence of polymers reduces the iron nanoparticle cluster sizes in the nanocomposites. As a result of this, interparticle exchange interactions are reduced due to increased interparticle separations, thereby further reducing the magnetization. Thus with increasing polymer concentration the system is dominated by anisotropy and dipolar interactions typical to all single domain magnetic nanoparticle systems.

The magnetization was also recorded for water-based samples in the liquid form for two samples namely, pure iron nanoparticles and 3:1 (Fe: PVP) sample and the results are shown in figure 6. The saturation magnetization values are comparable with those of the same samples in the powder form. The saturation magnetization ($M_s$) values obtained for the water-based iron nanoparticles in the presence and absence of PVP were 25 emu / gm and 55 emu / gm, respectively.

Assuming lognormal size distributions, in the perfectly non-interacting picture it is possible to estimate theoretically the average size of the iron particles from the room temperature magnetization data, using the method described by Chantrell et al (15). Using these results the average size of the iron nanoparticles for the magnetization data reported in Fig. 4 is ~ 4.45 nm for the pure iron nanoparticles and ranges from ~3.25 nm for the 3:4 sample and ~4.15 nm for the 3:1 sample. For the magnetization data of Fig 6, the calculated sizes are 4.2 nm for the PVP stabilized iron nanoparticles and 4.6 nm for the non-stabilized particles. However, this is just an estimate of the size of the magnetic core and cannot tell us anything about the extent of the non-magnetic oxide coating or the thickness of the PVP layer.

### IV. Monte Carlo simulations

In order to explain the trends observed in the magnetization results (Figures 4, 6 and 7) where the reduction in magnetization with increasing proportion of polymer in the compound cannot be fully accounted for by mass effects, we perform numerical simulations. The simulation system comprises of an array of single domain magnetic particles with varying sizes distributed at random within a simulation cell and interacting via exchange and dipolar interactions. The Hamiltonian for such a system can be written as follows,

$$H = -K \sum_i V_i \frac{(\vec{m}_i \cdot \vec{n}_i)^2}{|\vec{m}_i|^2} - \sum_{\langle i \neq j \rangle} J_{ij} \vec{m}_i \cdot \vec{m}_j - m_0 \sum_{\langle i \neq j \rangle} \frac{3(\vec{m}_i \cdot \vec{e}_{ij})(\vec{m}_j \cdot \vec{e}_{ij}) - \vec{m}_i \cdot \vec{m}_j}{r_{ij}^3} - m_0 \sum_j \vec{H} \cdot \vec{m}_j$$

Where, the first, second, third and fourth terms represent the anisotropy, exchange, dipolar and Zeeman field energies for the system respectively. $K$ is the anisotropy constant for the system and $V_i$, $\vec{m}_i = M_s V_i \vec{S}_i$ and $\vec{n}_i$ represent the volume, magnetization and unit vector along the easy axis direction of the $i$ th particle respectively, and $M_s$ is the saturation magnetization of the bulk system, $J_{ij}$ is the ferromagnetic exchange interaction between the $i$ th and $j$ th particles and $\vec{r}_{ij}$ is the distance between the same, $\vec{e}_{ij}$ being the unit vector along $\vec{r}_{ij}$. $\vec{H}$ is the externally applied magnetic field. The particle volumes are picked from a normal distribution using a mean size of a sphere of diameter 5 nm and the width of the distribution is (1). The parameters used for the simulation are: the anisotropy energy of the bulk system = 53 kJ/m$^3$, the saturation magnetization as 1750 k A/m and effective exchange interaction energy ~ 0.2 times the average anisotropy energy.

The simulation for the hystersis of this system of particles is carried out at room temperature using the standard Metropolis algorithm. The simulation system and other details are described elsewhere [16, 17]. The results of the simulation are presented in Figure 8. In this figure curve (a) represents the hysteresis for a system of strongly clustered fine Fe particles, with strong, dominant intra-cluster exchange interactions and relatively weak dipolar interactions due to large inter-cluster distances. This represents the physical picture corresponding to the TEM of Fig 2(a), where one has fairly large, well-separated iron clusters in water. On the other hand curve (b) represents the simulated hysteresis for a physical system where the large iron clusters have broken down into much smaller ones with smaller inter-cluster distances. This then is a system where dipolar interactions dominate over the exchange interactions, as would be the case seen in Fig 2(b). In keeping with experimental results we can see that in (b) the simulated magnetization is less than in (a) which is an effect of stronger dipolar interactions in (b).

Thus it is expected that as the proportion of polymer in the material increases accompanied by the breaking down of clusters and a more uniform distribution of these smaller clusters, a reduced magnetization would be seen in the system due to the gradual dominance of inter cluster dipolar interactions over the intra-cluster exchange interactions.

### 4. Conclusions

The iron nanoparticles obtained by the electro exploding techniques exhibit very interesting optical properties. Fluorescence was observed in the visible region for the pure iron particles and an enhancement in the signal was observed for the PVP coated nanoparticles. The magnetic properties of the iron nanoparticles and the PVP coated nanoparticles have been explained using an array of single domain exhibiting dipolar interaction. The iron nanoparticles in water medium can find applications in sensing water contamination.

**Acknowledgments**

We would like to acknowledge Dr. N C Mehra and Mr. Raman of University Science Instrumentation Center (USIC), and Mr. P C Padmakshan of the Geology Department, University of Delhi for recording TEM, UV-Visible, and XRD respectively, Dr. N K Chaudhary of Institute for Nuclear Medicines and Allied Sciences (INMAS), Delhi for recording the fluorescence spectra. We would also like to acknowledge the Department of Science and Technology (DST), India for the funding through the project (SR/S5/NM-52/2002) from the Nanoscience and Technology Initiative programme.

**Figure captions:**

**Figure 1**. X-ray diffraction for iron nanoparticles with respect to bulk iron wire used in the explosions

**Figure 2.** Transmission electron microscopy and size histogram for (a) pure iron nanoparticles and (b) PVP coated iron nanoparticles. Inset in (a) shows the electron diffraction patterns of pure iron nanoparticles

**Figure 3**. (a) UV-Visible spectra and (b) Fluorescence excitation and emission spectra, for both pure iron nanoparticles and PVP coated iron nanoparticles

**Figure 4**. Experimental hysteresis at 300 K for powder of pure iron nanoparticles and three different compositions of Fe -PVP nanoparticles which are 3:1, 3:2 and 3:4 (Fe: PVP)

**Figure 5**. Thermal analysis for the composition 3:1 (Fe: PVP) PVP coated iron nanoparticles with respect to pure PVP

**Figure 6**. Experimental hysteresis at 300 K for water phase iron nanoparticles and 3:1 (Fe: PVP) PVP coated iron nanoparticles

**Figure 7.** Weight corrected magnetizations of 3:1 (Fe: PVP) with respect to pure iron nanoparticles for (a) powder samples and (b) liquid samples

**Figure 8**. Simulated hysteresis at 300 K (a) represents the hysteresis for a system of strongly clustered (~60 -70 nm) fine Fe particles, with strong, dominant intra-cluster exchange interactions and relatively weak dipolar interactions (b) represents the simulated hysteresis for a physical system where the large iron clusters have broken down into much smaller ones (~ 10 nm) with smaller inter-cluster distances

**Abdullah Alqudami**

**Magnetic properties of iron nanoparticles prepared by exploding wire technique**

Figure 1.

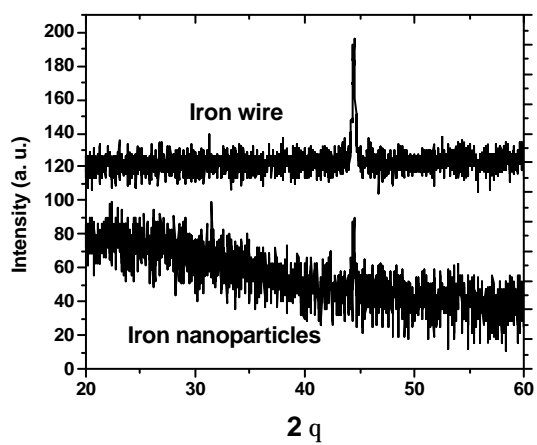

**Abdullah Alqudami**

**Magnetic properties of iron nanoparticles prepared by exploding wire technique**

Figure 2.

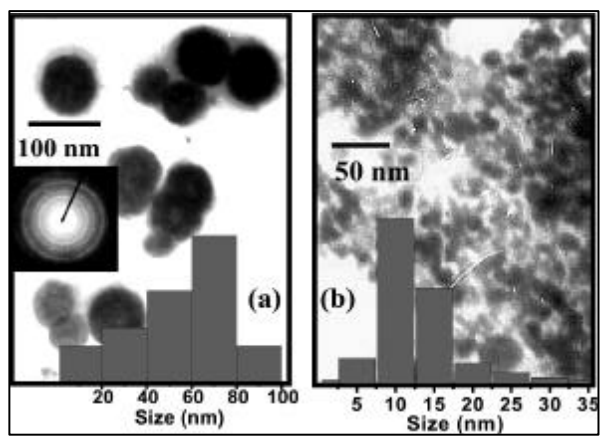

**Abdullah Alqudami**

**Magnetic properties of iron nanoparticles prepared by exploding wire technique**

Figure 3.

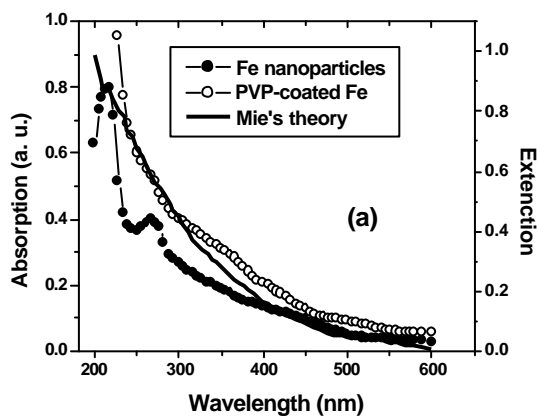

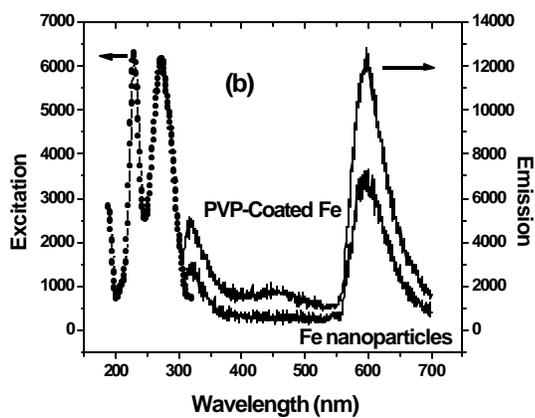

**Abdullah Alqudami**

**Magnetic properties of iron nanoparticles prepared by exploding wire technique**

Figure 4.

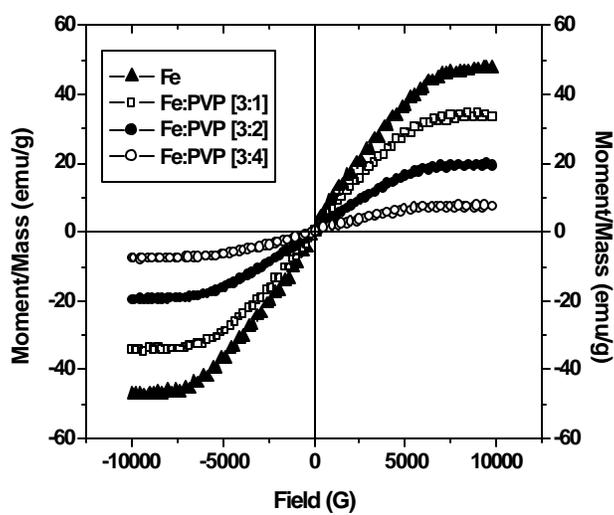

# Abdullah Alqudami

# Magnetic properties of iron nanoparticles prepared by exploding wire technique

Figure 5.

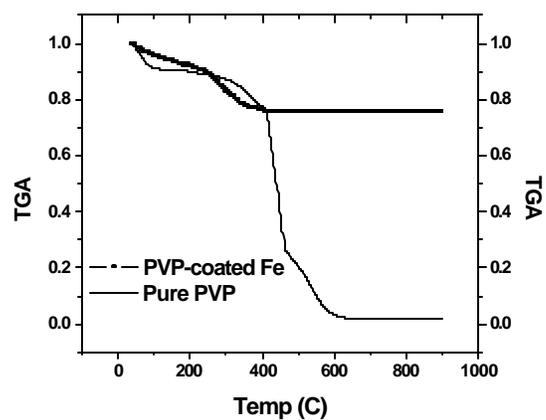

**Abdullah Alqudami**

**Magnetic properties of iron nanoparticles prepared by exploding wire technique**

Figure 6.

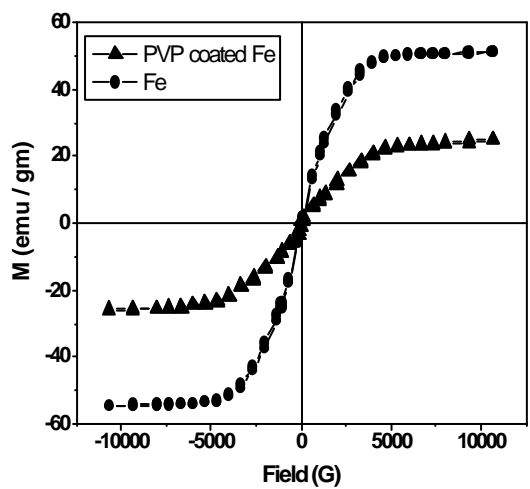

**Abdullah Alqudami**

**Magnetic properties of iron nanoparticles prepared by exploding wire technique**

Figure 7.

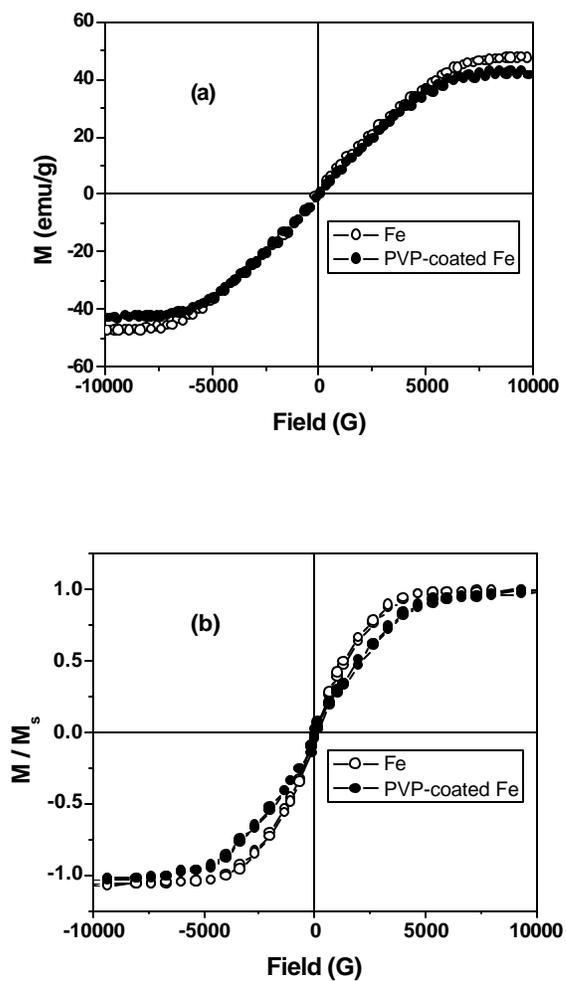

**Abdullah Alqudami**

**Magnetic properties of iron nanoparticles prepared by exploding wire technique**

Figure 8.

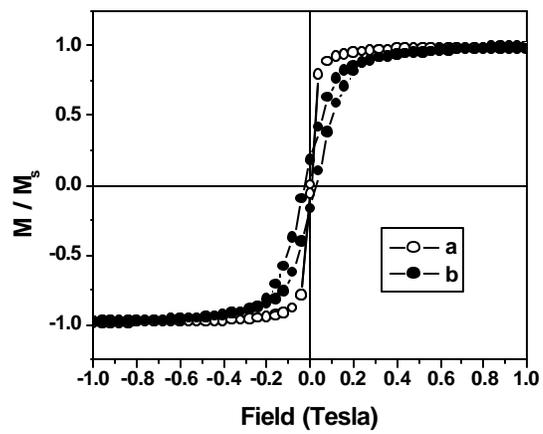